\def\bk{{\bf k}}
\def\br{{\bf r}}
\def\w{\omega}
\def\wk{\omega_k}
\def\w0{\omega_0}
\def\epkj{\epsilon_{\bk j}}
\def\ekj{\hat{e}_{\bk j}}

\documentclass[preprint,aps,showpacs]{revtex4}

\begin{document}

\title{Time-dependent Casimir-Polder forces \\
and  partially dressed states}

\author{R. Passante}
\affiliation{Istituto di BioFisica - Sezione di Palermo, 
Consiglio Nazionale delle Ricerche, Via Ugo La Malfa 153, I-90146 Palermo, Italy}

\email{passante@iaif.pa.cnr.it}
\author{F. Persico}

\affiliation{INFM and Dipartimento di Scienze Fisiche ed Astronomiche, 
Universit\'{a} degli Studi di Palermo, Via Archirafi 36, I-90123 Palermo, Italy}

\date{\today}

\begin{abstract}
A time-dependent Casimir-Polder force is shown to arise during the time
evolution of a partially dressed two-level atom. The partially dressed atom is obtained
by a rapid change of an atomic parameter such as its transition frequency, due
to the action of some external agent. The electromagnetic field fluctuations around
the atom, averaged over the solid angle for simplicity, 
are calculated as a function of time, and it is shown that the interaction
energy with a second atom yields a dynamical Casimir-Polder potential between
the two atoms.
\end{abstract}


\maketitle

Time-dependent phenomena in quantum electrodynamics have
recently been the object of a renewed interest, particularly in
connection with the effects which involve time-dependent
quantum fluctuations of the vacuum and lead to
time-dependent Casimir forces (see e.g. \cite{Dodonov98}).
On the other hand,
Casimir-Polder intermolecular interactions are long-range 
interactions between neutral atoms or molecules, originating from
their common interaction with the (vacuum) quantum
radiation field \cite{CP48,Milonni95}. Like the Casimir forces, 
they are also a manifestation
of the quantum nature of the electromagnetic field.
In the case of two isotropic ground-state atoms, the Casimir-Polder potential
behaves as $r^{-6}$ in the so-called ``near zone'' and as $r^{-7}$ in the ``far zone'', 
$r$ being the intermolecular separation, assumed larger than the 
overlap region of the electronic wavefunctions of the atoms. The length scale
separating the near and far zones is a typical (average) atomic transition
wavelength from the ground state \cite{CPP95}. The near-zone part of the
potential is the usual van der Waals potential for neutral atoms or
polarizable bodies, and involves only the electrostatic dipole-dipole
interaction between the fluctuating atomic dipole moments. 
The far zone part of the potential involves also the interaction
with the transverse radiation
field, and in particular it stems from 
the exchange of pairs of virtual photons between the atoms.

Many physical models for the Casimir-Polder forces
have been proposed, in order to gain a qualitative
understanding of the origin of these intermolecular forces. 
Some of the models
proposed trace the physical origin of the Casimir-Polder potential to the
vacuum fluctuations and some others to 
the radiation reaction field (for a review see \cite{CPP95} and 
references therein;
three- and many-body Casimir-Polder forces have also been considered
in the literature\cite{AZ60,PT85,CP97}).
The Casimir-Polder potential between ground-state neutral 
atoms has also been related to the presence of a virtual photon cloud around
each ground-state atom, both for the two- \cite{PP87} 
and many-body cases \cite{PP99}.
In fact, the virtual photons dressing an atom generate space-dependent
fluctuations of the electric and 
of the magnetic field around the atom, and their 
interaction with other atoms yields the Casimir-Polder potential
\cite{CPP95}. 
Thus Casimir-Polder forces 
probe the field fluctuations generated by an atom in vacuo. 
In other words, the Casimir-Polder potential between
two atoms directly yields the electromagnetic field fluctuations generated
by the virtual photon cloud dressing a ground-state neutral atom.
More precisely, it is the average force which is observed, because
Casimir-Polder forces are fluctuating forces \cite{WKF02}.

The situation becomes more interesting if one of the atoms is in a dynamic
situation, because a change in time of the electromagnetic field fluctuations
is expected in this case, leading to time-dependent Casimir-Polder forces. 
This was investigated in the case of the
dressing of an initially bare atom and in the complementary case of the
undressing of a fully dressed atom
using a simple effective Hamiltonian for the atom-field interaction \cite{CPP88}. 
In the first case, it was found that: 
i) at time $t$, the electric and magnetic energy densities of the field fluctuations 
around the dressing atom are the same as those of the fully dressed
source for $r<ct$, while for $r>ct$ they are the same as in 
the absence of the atom, as required
by the causality principle; ii) at $r=ct$, a singularity of the energy density is 
present, which propagates with velocity $c$; iii) the interaction energy of these
field fluctuations with the second ``test" atom 
yields a time-dependent potential \cite{CPP88}.
This potential is the analogue of the Casimir-Polder potential in a
dynamical situation. 
In the second case complementary results were obtained.
A bare field source however cannot  be easily produced in the
laboratory. Thus the evolution from a bare to a
dressed state, which was considered in \cite{CPP88}, 
is an extreme situation which cannot be considered realistic because the
atom-field interaction can never be switched off. 
In addition, the evolution from bare to dressed
states significantly involves high-frequency modes of the field, which are
known to generate the ultraviolet divergences of quantum field theory.
For these reasons, we have more recently considered 
a more sophisticated model where the evolution of an initially
partially dressed atom is involved \cite{PV96}. Partially 
dressed states of an atom are dressed states 
with an incomplete virtual photon cloud \cite{Fe66}. This model 
is free from the unrealistic assumption of an 
initially bare atom, and in \cite{PV96} 
it has been shown that the high-frequency modes of the radiation field, 
which would eventually lead to the
ultraviolet divergences, do not play a significant role in the evolution of
the system. 

In this letter we consider the time-dependent
energy density of the electromagnetic field
fluctuations around a partially dressed atom, and calculate explicitly the
dynamics of the electromagnetic field fluctuations around it,
averaging over the solid angle for simplicity. We also
consider their interaction energy with a second 
``test'' atom and the resulting time-dependent
Casimir-Polder force.
We use the same model of  partially dressed atom as  in
\cite{PV96}; the reader is referred to this paper for more detail on the model used.

We consider a two-level atom interacting with the electromagnetic
radiation field, in the Coulomb gauge and within dipole approximation. The
Hamiltonian describing this system is ($\hbar = 1$)
\begin{equation}
H' = \w0 S_z + \sum_{\bk j} \wk a_{\bk j}^\dagger a_{\bk j}
	+\sum_{\bk j} \epkj \left( a_{\bk j} - a_{\bk j}^\dagger \right)
	\left( S_+ - S_- \right)
\label{eq:1}
\end{equation}
$S$ are the pseudospin atomic operators and
the coupling constant $\epkj$, in the multipolar coupling scheme is \cite{CPP95}
\begin{equation}
\epkj = -\sqrt{\frac{2\pi \wk}V} \ekj \cdot {\bf d}
\label{eq:2}
\end{equation}
where ${\bf d}$ is the atomic dipole moment.

We assume that up to time $t=0$ the atom is in its dressed ground state.
The dressed ground state, at second-order in the electric charge, is
\begin{eqnarray}
\mid g \rangle &=& \left( 1 -\frac 12 \sum_{\bk j} 
\frac {\epkj^2}{(\w0 + \wk)^2} \right) \mid \downarrow 0_{\bk j} \rangle
+ \sum_{\bk j} \frac \epkj{\w0 + \wk} \mid \uparrow \bk j \rangle
\nonumber \\
&-& \sum_{\bk j {\bf k'} j'} \frac {\epkj \epsilon_{{\bf k'}j'}}
{(\w0 +\wk)(\wk +\omega_{k'})}
\mid \downarrow \bk j {\bf k'} j' \rangle
\label{eq:3}
\end{eqnarray}
where $\downarrow (\uparrow)$ denotes the atomic ground (excited) state,
and $\bk j$ indicates photon states.

We now assume that at $t=0$ the atom is subjected to an abrupt change 
$\Delta \w0$ in its transition frequency (for example, by
the action of an external electric field; see \cite{PV96} for more detail). 
Abrupt changes in the parameters of atom-photon Hamiltonian have
been considered previously in connection with radiative processes of 
different kinds (see e.g. \cite{TSJ96}).
In our case, the atomic
transition frequency is $\w0$ for $t<0$ and $\w0 + \Delta \w0$ for $t>0$.
A consequence of this change is that the state (\ref{eq:3}), which
is an eigenstate of $H'$ for $t<0$, is not an
eigenstate of the total Hamiltonian for $t>0$, because the Hamiltonian
has changed. Thus, the state $\mid g \rangle$  
must evolve for $t>0$, and in this way we have generated a partially
dressed state. The Hamiltonian of the system for $t>0$ is then
\begin{equation}
H=H' +\Delta \w0 S_z
\label{eq:4}
\end{equation}
where $H'$ is the Hamiltonian for $t<0$, given by (\ref{eq:1}).

The state of the system at $t>0$ has been obtained in \cite{PV96} at the second order
in $\epkj$ and first order in $\Delta \w0$. Using this state, we can evaluate
at $t>0$ the change of the electric and magnetic energy density around the atom
with respect to that at times $t<0$. After lengthy calculations, we obtain ($t>0$)
\begin{eqnarray}
\delta {\cal E}_E (\br ,t) &=& \frac 1{8\pi}  \langle E^2(\br ,t) \rangle 
- \frac 1{8\pi} \langle  E^2(\br , t<0) \rangle
\nonumber \\
&=& -\Delta \w0 \frac {c^2}{(2\pi )^3} \int \int dk dk' k^3 k'^3
\left\{ \left( 
d^2 j_0(kr) j_0(k'r) + \frac 1{k'^2} j_0(kr) \left( {\bf d} \cdot \nabla^r \right)^2 j_0(k'r)
\nonumber \right. \right. \\
&+& \left.  \frac 1{k^2} j_0(k'r) \left( {\bf d} \cdot \nabla^r \right)^2 j_0(kr) 
+ \frac 1{k^2k'^2} d_i d_j \left( \nabla_\ell^r \nabla_i^r j_0(kr) \right)
\left( \nabla_j^r \nabla_\ell^r  j_0(k'r) \right) \right)
\nonumber \\
&\times& \left.
F(\wk, \omega_{k'},t) \right\} +cc
\label{eq:5}
\end{eqnarray}
for the electric energy density, where
$\nabla^r$ is the gradient operator with respect to the
coordinate $\br$, and
\begin{eqnarray}
\delta {\cal E}_M (\br ,t) &=& \frac 1{8\pi} \langle B^2(\br ,t) \rangle 
- \frac 1{8\pi} \langle B^2(\br , t<0) \rangle
\nonumber \\
&=& \Delta \w0 \frac {c^2}{(2\pi )^3} \int \int dk dk' k^2 k'^2
\left\{ \left( 
d^2 \nabla^r(j_0(kr)) \cdot \nabla^r (j_0(k'r)) \right. \right.
\nonumber \\ 
&-& \left. \left. ( {\bf d} \cdot \nabla^r j_0(kr) ) ( {\bf d} \cdot \nabla^r j_0(k'r) )
\right) 
G(\wk, \omega_{k'},t) \right\} +cc
\label{eq:6}
\end{eqnarray}
for the magnetic energy density. The functions $F(\wk, \omega_{k'},t)$ and
$G(\wk, \omega_{k'},t)$ assume a simple form in the far zone
$r>>c/\w0$, where we can assume $\wk << \w0$ \cite{CT98}.
In this case, we have
\begin{equation}
F(\wk, \omega_{k'},t) = G(\wk, \omega_{k'},t)
= \frac 1{\w0^2(\wk + \omega_{k'})} \left( 1
-e^{-i(\wk + \omega_{k'})t} \right)
\label{eq:7}
\end{equation}

In order to simplify the calculation of (\ref{eq:5},\ref{eq:6}), we perform
an average of the energy densities over a sphere of radius $r$ centered
on the atom. The integrals on $k,k'$ can be performed explicitly, and
the final result is (valid only in the far zone)
\begin{eqnarray}
\bar{\delta {\cal E}}_E (r ,t) &=& \frac 1{4\pi} \int d\Omega 
{\delta {\cal E}}_E (\br ,t) 
\nonumber \\
&=& -\Delta \w0 
\frac {cd^2}{24\pi^2 \w0^2} \left( \frac {13}{2r^7} \left( 1 -\Theta (r-ct) \right)
+\frac {13}{2r^6} \delta (r-ct) - \frac 5{2r^5} \delta '(r-ct) \right.
\nonumber \\
&+& \left. \frac 1{3r^4}
\delta ''(r-ct) + \frac 1{30r^2} \delta^{(iv)} (r-ct) \right)
\label{eq:8}
\end{eqnarray} 

\begin{eqnarray}
\bar{\delta {\cal E}}_M (r ,t) &=& \frac 1{4\pi} \int d\Omega 
{\delta {\cal E}}_M (\br ,t) 
\nonumber \\
&=& \Delta \w0 
\frac {cd^2}{24\pi^2 \w0^2} \left( \frac {133}{2r^7} \left( 1 -\Theta (r-ct) \right)
+\frac {133}{2r^6} \delta (r-ct) - \frac {29}{r^5} \delta '(r-ct) \right.
\nonumber \\
&+& \left. \frac {41}{6r^4} \delta ''(r-ct) - \frac 5{6r^3} \delta '''(r-ct)
+\frac 1{30r^2} \delta^{iv} (r-ct) \right)
\label{eq:9}
\end{eqnarray} 

These expressions have a structure similar to that obtained for the
dressing of the bare source \cite{CPP88}. We note that the
most diverging terms, i.e. those proportional to the fourth 
derivative of the delta function, 
cancel in the total (electric plus magnetic) energy density,
coherently with the known fact the the 
time evolution of the field energy in this model
is less affected by the high-frequency field modes compared to the
bare atom case \cite{PV96}.
Moreover, they are proportional to the atom's electric static polarizability
$\alpha \sim d^2/\w0$. The further proportionality to
$\Delta \w0 /\w0$, which is a quantity much smaller than one, indicates
that all single-atom effects related to the time-dependent field 
fluctuations given by
(\ref{eq:8}) and (\ref{eq:9}) are quite small, although in a system of 
many atoms the effects may become measurable.

The presence of the $\Theta$ function in (\ref{eq:8},\ref{eq:9}) ensures causality
in the propagation of the energy density. A singularity of the
electric and magnetic energy density is present at $r=ct$. 
When a second atom, described
by its static electric polarizability $\alpha$, 
assumed isotropic for simplicity, is placed at a distance
$r$ from the first atom, the resulting Casimir-Polder interaction potential between
the two atoms can be obtained as
\begin{equation}
V(\br ,t) = -\frac 12 \alpha \langle E^2(\br ,t) \rangle
\label{eq:10}
\end{equation}
where $\alpha$ is the static polarizability of the second atom \cite{CP69,PPT98}.
A similar interaction energy involving the magnetic field fluctuations exists, proportional to
the magnetic polarizability of the second atom. The potential energy
$V(\br ,t)$ in
equation (\ref{eq:10}) yields a time-dependent Casimir-Polder potential between
the two atoms; contrarily to previous calculations, it has been calculated in a 
situation which does not involve unrealistic bare states, because it
results from the evolution of a partially dressed atom. 
This result is the conterpart of the dynamical Casimir effect; 
however, a noteworthy difference is
that in the present case matter has been treated dynamically, 
whereas the usual description
of the dynamical Casimir effect includes the presence of matter only 
through the boundary condition 
on the field operators \cite{BMM01}; only very recently, 
corrections to the results based on external boundary conditions have been
introduced in the calculation of the Casimir effect for
macroscopic bodies for a 1D cavity with a moving mirror \cite{SH02}. 
As mentioned in the introduction, the main point of this paper is to show that a
time-dependent Casimir-Polder counterpart of the time-dependent Casimir force
exists in QED and does not depend on unrealistic assumptions such as an
initially bare atom.
A reliable estimate of the intensity of this force requires a more detailed
model for the partially dressed atom, but it is 
evident that the dynamical Casimir-Polder interaction investigated 
in this paper within an isolated pair of atoms, 
although conceptually important, is a very tiny effect. In fact,
the time-dependent force, compared
to the usual time-independent Casimir-Polder force, which has been directly
observed \cite{SBCSH}, contains an extra factor $\Delta \w0 /\w0 << 1$.
We hope to discuss this point in a future publication.

In conclusion, we have considered a model of a partially dressed field source
in the framework of quantum electrodynamics. 
This model consists of a dressed two-level atom, 
interacting with the electromagnetic radiation field, which is suddenly subjected to
a rapid change of its transition frequency due to an external agent.  The dressed
ground state of the atom becomes a partially dressed state after this change.
The time evolution of the energy density of its dressing photon cloud 
has been considered, and it has been shown that this process yields 
a time-dependent Casimir-Polder force on a second
neutral atom placed at a distance $r$. 
We have shown that this new effect is in principle detectable, 
since the treatment we have given is free of the unrealistic
assumption of an initial bare atom.

This work was supported by the European Union under contract No. HPHA-CT-2001-40002
and in part by the bilateral Italian-Japanese project 15C1 on Quantum
Information and Computation of the Italian Ministry for Foreign Affairs. 
Partial support by Ministero dell'Universit\'{a} e della Ricerca Scientifica 
e Tecnologica and by Comitato Regionale di Ricerche Nucleari e di Struttura della Materia
is also acknowledged.

\end{document}